\documentclass[sigconf]{acmart}

\usepackage{booktabs} 
\usepackage{color}

\usepackage{caption}
\usepackage{times}
\usepackage{epsfig}
\usepackage{graphicx}
\usepackage{amsmath}
\usepackage{amsfonts}
\usepackage{amssymb}
\usepackage{calligra}
\usepackage{calrsfs}
\usepackage{mathtools}
\usepackage{url}
\usepackage{booktabs}
\usepackage{multirow}
\usepackage{multicol}
\usepackage{siunitx}

\acmDOI{}

\acmISBN{}

\acmConference[SysML2018]{ACM SysML conference}{February 2018}{California, USA}
\acmYear{2018}

\begin{document}
\title{Corpus Conversion Service: \\A machine learning platform to ingest documents at scale.}

\author{Peter W J Staar, Michele Dolfi, Christoph Auer, Costas Bekas}
\email{ taa, dol, cau, bek @zurich.ibm.com}
\affiliation{%
  \institution{IBM Research}
  \streetaddress{14 Saumerstrasse}
  \city{Rueschlikon}
   \country{Switzerland}
}

\setcopyright{none}

\maketitle

\section{\label{sec:Introduction}Introduction}

PDF is by far the most prevalent document format today. There are roughly 2.5 trillion PDFs in circulation~\cite{billionsPdfs} such as scientific publications, manuals, reports, contracts and more. However, content encoded in PDF is by its nature reduced to streams of printing instructions purposed to faithfully present a visual layout. The task of automatic content reconstruction and conversion of PDF documents into structured data files has been an outstanding problem for over three decades  ~\cite{Cattoni98geometriclayout, Chanod2005}. Here, we present a solution to the problem of document conversion, which at its core uses trainable, machine learning algorithms. The central idea is that we avoid heuristic or rule-based (RB) conversion algorithms, using instead generic machine learning (ML) algorithms, which produce models based on gathered ground-truth data. In this way, we eliminate the continuous tweaking of conversion rules and let the solution simply learn how to correctly convert documents by providing enough ground truth. This approach is in stark contrast to current state of the art conversion systems (both open-source and proprietary), which are all RB.

While a machine learning approach might appear very natural in the current era of AI, it has serious consequences with regard to the design of such a solution. First, one should think at the level of a document collection (or a corpus of documents) as opposed to individual documents, since an ML model for a single document is not very useful. An ML model for a certain type of documents (e.g. scientific articles, regulations, contracts, etc.) obviously is. Secondly, one needs efficient tools to gather ground truth via human annotation. These annotations can then be used to train the ML models. It is clear then that leveraging ML adds an extra level of complexity: One has to provide the ability to store a collection of documents, annotate these documents, store the annotations, train models and ultimately apply these models on unseen documents. For the authors of this paper, this implied that our solution cannot be a monolithic application. Rather it was built as a cloud-based platform, which consists out of micro-services that execute the previously mentioned tasks in an efficient and scalable way. We call this platform \textit{Corpus Conversion Service} (CCS).

\begin{figure}[t]
\center
\includegraphics[scale=0.5]{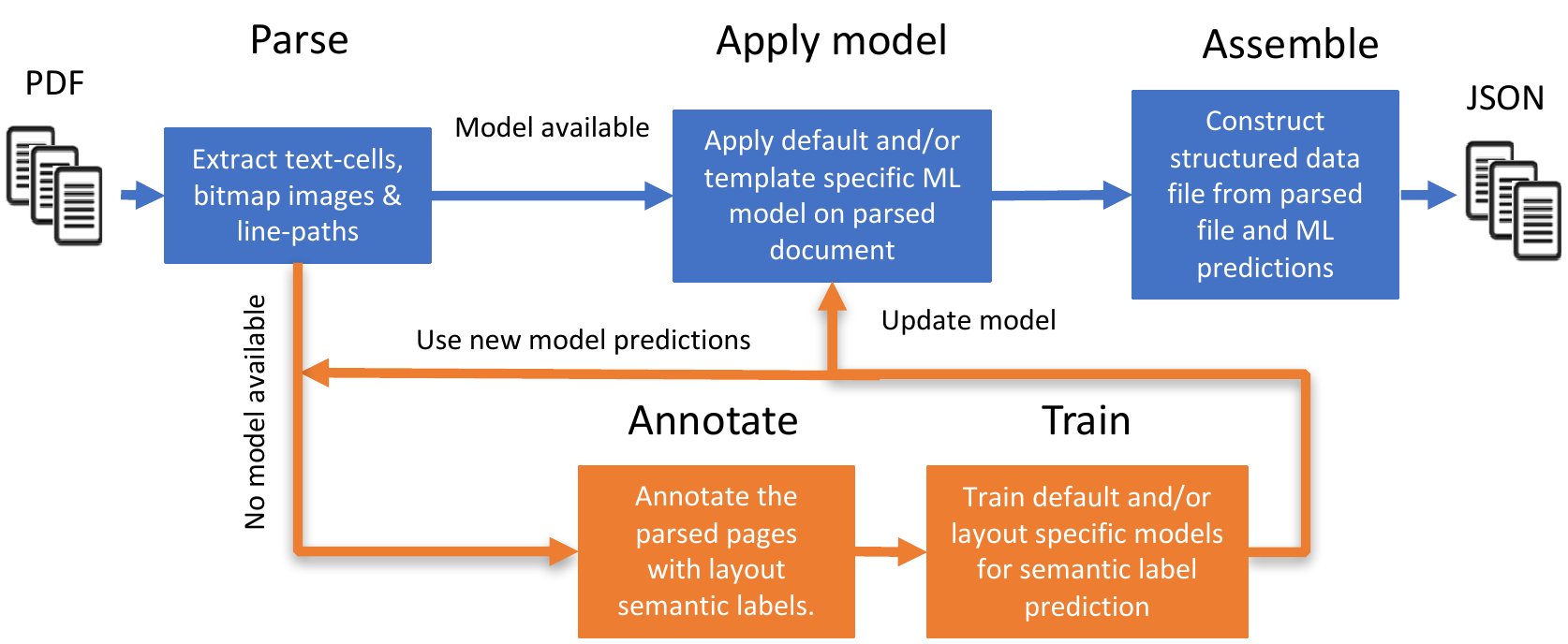}
\caption{\label{fig:SApipeline} A sketch of the \textit{Corpus Conversion Service} platform for document conversion. The main conversion pipeline is depicted in blue and allows you to process and convert documents at scale into a structured data format. The orange section can be used optionally, in order to train new models based on human annotation. }
\end{figure}

\section{\label{sec:Method}Platform and its microservices}

\begin{figure}[b]
\center
\includegraphics[width=3.5in]{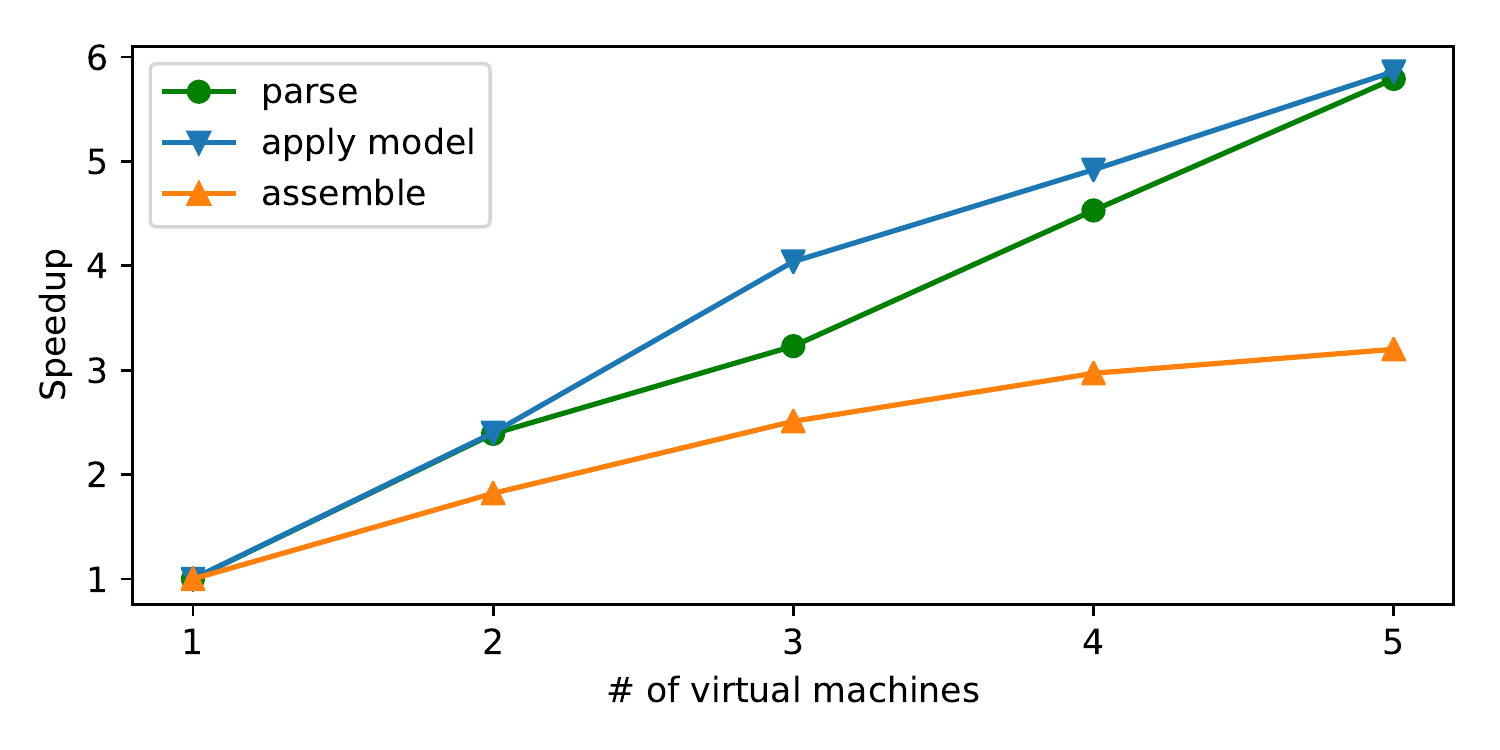}
\caption{\label{fig:ScalingResources} Number of processed pages per second as a function of the number of virtual machines (4 core, 8 Gigabytes RAM).}
\end{figure}

Using a micro-services architecture, the CCS platform implements a pipeline on the cloud.  
This pipeline consists out of micro-services, which can be grouped into five components. These are (1) the parsing of documents into an internal format optimised for ML, (2) applying the ML model(s), (3) assembling the document(s) into a structured data format, (4) annotating the parsed documents and (5) training the models from the annotations. These components are shown in Figure~\ref{fig:SApipeline}.
If a trained model is available, only three of the components are needed to convert the documents, namely parsing, applying the model and assembling the document. As is shown in Figure~\ref{fig:ScalingResources}, the microservices in these components are designed to scale with regard to compute resources (i.e. virtual machines) in order to keep time-to-solution constant, independent of the load. 
If no machine learned model is available, we have two additional components, i.e annotation and training, that allow us to gather ground truth and train new models. 

\begin{figure}[!t]
\center
\frame{\includegraphics[width=3.2in]{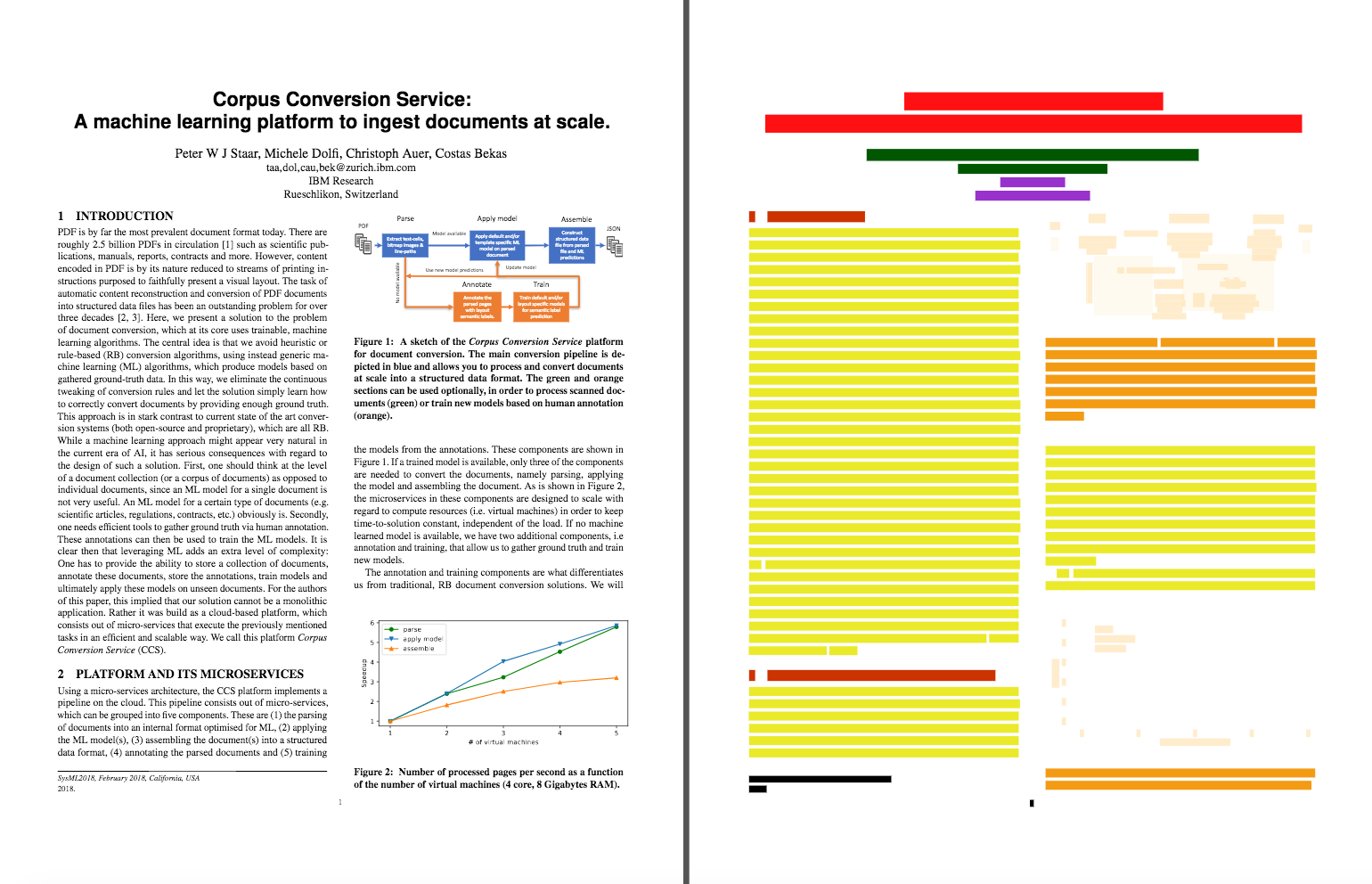}}
\caption{\label{fig:SAAnnotation} The annotated cells obtained for the title page of this document. Here, the \textit{title}, \textit{authors}, \textit{affiliation}, \textit{subtitle}, \textit{main-text}, \textit{caption} and \textit{picture} labels are represented respectively as red, green, purple, dark-red, yellow, orange and ivory.}
\vspace{-1em}
\end{figure}

The annotation and training components are what differentiates us from traditional, RB document conversion solutions. We will therefore focus primarily on those components in the remainder of this document.
In the parsing phase of the pipeline, we are focused on the following straightforward but non-trivial task: \emph{Find the bounding boxes of all text-snippets that appear on each pdf-page.} For simplicity, we will refer to the bounding boxes of the text-snippets as \textit{cells}. 
In Figure~\ref{fig:SAAnnotation}, we show the cells obtained from the title-page of this paper. On the CCS platform, we consider the cells (i.e a bounding box with their associated text) as the atomic pieces of content from the original document. The goal of the ML algorithms is now to associate with these cells certain classes (e.g. layout semantics), that allow us to recombine these cells into a structured data file which contains the content of the original document. For text-reconstruction, it turns out that simply predicting correctly the layout semantics (which we refer to as \textit{labels}) of each cell is sufficient. Examples of labels are: \textit{Title}, \textit{Abstract}, \textit{Authors}, \textit{Subtitle}, \textit{Text}, \textit{Table}, \textit{Figure}, etc\footnote{It is important to notice that there is no restriction on the number of labels nor their semantic meaning. The only limitation one has is that the set of semantic labels needs to be consistent across the dataset, but this is evidently true for any type of Machine Learning algorithm.}. 

In the annotation component, we have a page annotator that retrieves the original PDF page and its associated parsed components, containing the cells. We then ask the (human) annotator to assign each cell a label. In the page-annotator, each label is visually represented by a colour. By assigning a colour to each label, the task of annotation is translated into a \textit{colouring-task}, as can be seen in Figure~\ref{fig:SAAnnotation}. Since humans are very efficient in visual recognition, this task comes very natural to us. Consequently, the effort required for the tedious task of document-annotation is reduced significantly. This has been demonstrated numerous times during various annotation campaigns, where the average annotation time per document was reduced by at least one order of magnitude, corresponding to an ground-truth annotation rate of 30 pages per minute.

Once enough ground-truth has been collected, one can train ML models on the CCS platform. We have the ability to train two types of models: default models, which use state-of-the-art deep neural networks and customised models using Random Forest (RF) in combination with the default models. The aim of the default models is to detect objects on the page such as tables, images, formulas, etc. The customised ML models are classification models, which assign/predict for each cell on the page a label. In these customised models, we typically use the predictions of the default models as additional features to our annotation-derived features.

Due to the high variability across document layouts, we need very robust methods to detect these objects such as tables on a page. The most robust methods for detecting objects are currently deep neural networks for object-detection such as R-CNNs (and their derivatives Fast- and Faster-R-CNN)~\cite{GirshickDDM13:RCNN, Girshick:2015:FRCNN, NIPS2015_5638} and the YOLO architectures~\cite{Redmon2016YouOL, redmon2016yolo9000}. On our platform, we currently have the Faster-R-CNN~\cite{NIPS2015_5638} and the YOLOv2~\cite{ redmon2016yolo9000} networks available as individual micro-services, both for training and predictions. We have been able to train both networks to detect tables, on a dataset of 25000 pages originating from the arXiv.org repository. On a separate test-set of 5000 pages, we obtained for both of them recall and precision numbers above 97\%. 

The customised model training is designed to focus on a particular template, in order to boost the extraction accuracy. They are essentially RF models, which use the predictions of the default models in addition to the geometric features of the individual cells on the page. These geometric features of the cell can be directly obtained from the parsed document and are (1) the page-number, (2) the inverse page number (= total number of pages in the document - page-number), (3) the width and height of the cell and of the page, (4) the position of the cell ($=x_0, y_0, x_1, y_1$), (5) distance to the nearest neighbour cell (left, right, top and bottom), (6) the style of the text (normal, italic, or bold). In table~\ref{Table:PhysRevB}, we show the performance results for a customised model trained on the articles from the journal \textit{Physical Review B}\footnote{\url{https://journals.aps.org/prb}}. The latter follows a certain template layout. As one can observe, we obtain excellent recall and precision numbers, especially with regard to tables. This is not surprising, since we use the table detection of the default models in combination with an ML model that focuses on the template of the journal.

\begin{table}[H]
\centering
\caption{\label{Table:PhysRevB} Performance results for the template specific model of the \textit{Physical Review B} journals. The results were obtained via a 10-fold cross-validation procedure, in which we trained on 400 annotated pages.}
\begin{tabular}{l|l|llllllll||l||}
\toprule
                  & & \multicolumn{6}{c}{predicted label} \\ \hline
\multirow{8}{*}{\rotatebox[origin=c]{90}{true label}} &  & \rotatebox[origin=l]{90}{Title} & \rotatebox[origin=l]{90}{Author} & \rotatebox[origin=l]{90}{Subtitle\:} & \rotatebox[origin=l]{90}{Text} & \rotatebox[origin=l]{90}{Picture} & \rotatebox[origin=l]{90}{Table} \\ \hline 
                   & Title             & 75 & 0    & 0     & 0         & 0      & 0 \\
                  &  Author         & 1  & 670 & 0     & 0          & 0      & 0 \\
                  &  Subtitle        & 0  & 0     & 325 & 0          & 0      & 0 \\
                  &  Text             & 1  & 17      & 0    & 56460 & 14      & 0 \\
                  &  Picture        & 0  & 0      & 0     & 4        & 4223 & 26 \\
                  &  Table           & 0  & 0     & 0      & 0        & 1       & 3418   \\ \hline \hline 
                  & Recall& 100 &  99.85  & 100      & 99.94        & 99.24     & 99.97 \\
                  & Precision& 97.40 &   97.52  & 100      & 99.99        & 99.64      & 99.24 \\
\bottomrule
\end{tabular}
\end{table}

\bibliographystyle{unsrt}
\bibliography{acmart}

\end{document}